\documentclass[conference]{IEEEtran}
\IEEEoverridecommandlockouts
\usepackage{cite}
\usepackage{amsmath,amssymb,amsfonts}
\usepackage{algorithm}
\usepackage{algpseudocode} 
\usepackage{subfig}
\usepackage{graphicx}
\usepackage{textcomp}
\usepackage{xcolor}
\usepackage{balance}
\usepackage{url}
\usepackage{tabularx}
\usepackage{array}
\usepackage{booktabs}
\usepackage{siunitx}
\usepackage[
top=0.75in,
bottom=1.05in,
left=0.625in,
right=0.625in
]{geometry}

\def\BibTeX{{\rm B\kern-.05em{\sc i\kern-.025em b}\kern-.08em
    T\kern-.1667em\lower.7ex\hbox{E}\kern-.125emX}}
\begin{document}


\newcommand{\system}{\textit{POLARIS}} 

\title{POLARIS: PHY-Aware Spectrum Steering for Dynamic Spectrum Sharing
}

\author{\IEEEauthorblockN{Stavros Dimou}
\IEEEauthorblockA{\textit{Northeastern University}\\
Boston, MA, USA \\
dimou.s@northeastern.edu}
\and
\IEEEauthorblockN{Guevara Noubir}
\IEEEauthorblockA{\textit{Northeastern University}\\
Boston, MA, USA \\
g.noubir@northeastern.edu}
}

\IEEEaftertitletext{\vspace{-2.5\baselineskip}}

\maketitle

\begin{abstract}

Dynamic Spectrum Sharing (DSS) enables flexible activation of additional spectrum resources but leaves open a key runtime question: once new spectrum becomes available, which steering mechanism should migrate connected devices toward it with minimum service disruption?
We present the first PHY-aware characterization of 3GPP-compliant UE steering mechanisms, including Bandwidth Part (BWP) reconfiguration, Carrier Aggregation (CA), E-UTRA-NR Dual Connectivity (EN-DC), Connected-Mode Handover (HO), and Release and Redirection (R\&R), using modem-level traces from devices connected to operational networks, collected across 1,600 executions over four months in 12 urban areas.
By mapping each mechanism to observable PHY-layer milestones, we decompose steering latency into intrinsic PHY-centric execution and RRC-to-PHY completion components, revealing substantial heterogeneity: NR BWP achieves $6.25\,\mathrm{ms}$ mean latency with zero tail exceedance above $50\,\mathrm{ms}$, while CA exceeds $1225\,\mathrm{ms}$; mobility procedures remain largely modem-bound, whereas discovery-driven mechanisms experience significant RRC-to-PHY completion amplification.
Guided by these measurements, we design \system, an O-RAN-based system that selects the least disruptive steering mechanism via a two-parameter disruption score. 
\system\ reduces mean latency by up to $85.1\%$ and $T_{95}$ by $89.7\%$ over static or non-adaptive baselines, eliminates tail exceedance above $50\,\mathrm{ms}$, and avoids high-disruption mechanisms, demonstrating that PHY-layer execution profiling enables reliable and context-aware spectrum steering in DSS-enabled networks.

\end{abstract}


\section{Introduction}
\label{sec:intro}

The rapid growth of mobile connectivity is placing increasing pressure on radio spectrum utilization.  
Industry forecasts project several billion 5G subscriptions within the coming decade~\cite{EricssonMobility2025}, while spectrum (1–6 GHz) remains highly contested~\cite{SolidSpectrum}, and largely underutilized.
To address this inefficiency, regulatory initiatives such as localized spectrum licensing~\cite{ofcom} and the Citizens Broadband Radio Service (CBRS) framework~\cite{cbrs} have introduced mechanisms that promote more flexible shared access to spectrum resources. 
While these frameworks enable Dynamic Spectrum Sharing (DSS), they leave two 
questions largely unexplored: once spectrum resources become available, how should User Equipments (UEs) be steered toward them efficiently, and what performance cost do the available steering mechanisms impose on the UE?



\begin{table*}[t]
\centering
\caption{3GPP-compliant UE steering mechanisms for Dynamic Spectrum Sharing.}
\label{tab:steering_mechanism}
\scriptsize
\renewcommand{\arraystretch}{1}
\setlength{\tabcolsep}{4pt}

\begin{tabularx}{\textwidth}{|p{2.2cm}|p{2.3cm}|p{2cm}|X|p{1.6cm}|p{2.1cm}|}
\hline
\textbf{Mechanism} &
\textbf{Deployment Scenario} &
\textbf{UE State} &
\textbf{Operation Principle} &
\textbf{Synchronization Requirement} &
\textbf{Key 3GPP TS} \\
\hline

\textbf{BWP Reconfiguration} &
Intra-carrier, intra-frequency &
RRC\_CONNECTED &
gNB reconfigures UE to operate over an alternative BWP within the same carrier via \emph{RRCReconfiguration}. 
UE remains synchronized and applies updated bandwidth parameters. &
Serving Cell (synchronized) &
38.211; 38.213; 38.300; 38.331 \\ 
\hline

\textbf{Carrier Aggregation} &
Intra-/inter-frequency, intra-RAT &
RRC\_CONNECTED &
Additional spectrum is configured as a SCell. 
UE is reconfigured via \emph{RRCReconfiguration} 
while the Primary Cell (PCell) remains the control-plane anchor. &
SCell Sync &
38.300; 38.321; 38.331 \\
\hline

\textbf{EN-DC (NSA)} &
Inter-RAT (LTE-NR) &
RRC\_CONNECTED &
LTE Master Node configures an NR Secondary Node via \emph{RRCReconfiguration}. 
UE performs RACH toward the NR Primary Secondary Cell (PSCell) and establishes dual connectivity while LTE remains the control-plane anchor. &
NR PSCell Sync &
37.340; 38.331 \\
\hline

\textbf{Connected-Mode Handover} &
Intra-/inter-frequency, intra-RAT &
RRC\_CONNECTED &
Serving cell transfers UE to a target cell using \emph{RRCReconfiguration} with \emph{mobilityControlInfo}. 
UE synchronizes with the target, which becomes the new PCell. & 
Target Cell Sync &
38.300; 38.331 \\
\hline

\textbf{Release and Redirection} &
Intra-/inter-frequency, intra-/inter-RAT &
RRC\_CONNECTED $\rightarrow$ RRC\_IDLE &
Serving cell issues \emph{RRCRelease/RRCConnectionRelease} including \emph{redirectedCarrierInfo}. 
UE transitions to idle mode, performs reselection toward the target carrier, and re-establishes RRC connection. &
Target Cell Sync &
38.300; 38.304; 38.331 \\
\hline

\end{tabularx}
\end{table*}

3GPP specifies multiple mechanisms that can steer connected UEs toward newly available spectrum, including Bandwidth Part (BWP) reconfiguration, Carrier Aggregation (CA) Secondary Cell (SCell) addition, E-UTRA–NR Dual Connectivity (EN-DC), Connected-Mode Handover (HO), and Release and Redirection (R\&R). 
However, these mechanisms were originally designed for mobility, connectivity expansion, or load balancing rather than DSS. 
Although their signaling procedures are well specified in prior work, their behavior during spectrum activation, particularly their impact on UE-side Physical Layer (PHY-layer), remains largely unexplored.
As a result, a critical gap remains in understanding which steering mechanisms are most suitable for DSS and what UE-side disruption costs they introduce.

This gap becomes particularly relevant in the context of Open Radio Access Network (O-RAN) architecture, which introduces a new level of network programmability. 
By disaggregating the RAN into modular components, O-RAN replaces monolithic deployments with an open architecture.
Following principles similar to Software-Defined Networking (SDN), O-RAN supports network optimization through the deployment of RAN intelligent controllers (RICs). 
The RICs host modular “plug-and-play” applications, known as xApps and rApps, 
enabling advanced network optimization. 

To address these gaps, we present the first systematic characterization of 3GPP-compliant UE steering mechanisms in the context of DSS. 
Using real-world telemetry collected over four months across multiple U.S. locations from Commercial-Off-The-Shelf (COTS) UEs, we construct a novel mapping between steering procedures and UE PHY-layer milestones, enabling the decomposition of steering latency into PHY-centric and RRC-to-PHY completion components. 
Our analysis shows that BWP reconfiguration incurs the smallest PHY-centric disruption ($6.25\,\mathrm{ms}$), yet its total disruption increases to $1.97\,\mathrm{s}$ when accounting for RRC-to-PHY completion events.
We further observe that mobility procedures remain largely modem-bound, whereas discovery-driven mechanisms introduce significant modem reaction time, highlighting the distinct behavior of the modem across mechanisms.
Guided by these findings, we design \system, an O-RAN-based system that enables disruption-aware mechanism selection, reducing mean latency by up to  $85.1\%$ and $T_{95}$ by $89.7\%$ over static baselines, fully eliminating  tail exceedance above $50\,\mathrm{ms}$.


\section{Related Work}
\label{sec:related}

Recent work has explored DSS enabled by the programmability of O-RAN, introducing closed-loop spectrum management. 
Several studies investigate data-driven DSS within O-RAN, demonstrating how traffic patterns or spectrum occupancy can guide dynamic resource allocation~\cite{adapShare,Charm}. 
Other efforts examine O-RAN-assisted DSS in heterogeneous coexistence scenarios, 
enabled through centralized or distributed  coordination~\cite{qualcomm,Neutran,blockchain_enabled_net_sharing}. 
Collectively, these studies demonstrate the flexibility of O-RAN as a control substrate for DSS. 
However, they 
assume that connected UEs can seamlessly migrate to newly activated spectrum, abstracting away the steering procedures required to realize such transitions.

A substantial body of measurement research has also characterized operational cellular performance. 
Prior work investigates HO behavior and mobility management at scale~\cite{country_wide_handover,vivisecting_mobility}, analyzes CA performance~\cite{DissectingCA2024}, 
and evaluates the impact of BWP reconfiguration~\cite{impact_of_bwp_switching,primer_bwp}. 
Additional studies examine the performance differences between 5G Non-Standalone (NSA) and SA deployments, including the role of EN-DC across bands~\cite{nsa_performance,MidBandSigcomm2024}. 
While these works provide valuable insight into mechanisms,
they focus on high-level behavior, without exposing the underlying UE modem processing.
In contrast, this work presents the first unified UE-side PHY-layer characterization of 3GPP-compliant steering mechanisms.
By mapping steering procedures to observable UE PHY-layer milestones, we uncover cross-mechanism heterogeneity 
and leverage these insights to design \system.

\section{PHY-layer Steering and O-RAN Control}
\label{sec:background}

In this section, we provide the necessary background on the PHY-layer procedures involved in UE synchronization, operation of the steering mechanisms summarized in Table~\ref{tab:steering_mechanism}, and the O-RAN control framework that underpins our system.

\subsection{PHY-layer Synchronization}
\label{sec:phy_procedures}

UE steering manifests through PHY-layer procedures governing UE discovery, synchronization, and communication maintenance in both NR and LTE.
During cell acquisition, the UE detects synchronization signals, Synchronization Signal Blocks (SSBs) in NR or the Primary and Secondary Synchronization Signals (PSS/SSS) in LTE, to achieve time-frequency synchronization, followed by decoding of the Physical Broadcast Channel (PBCH) to obtain the Master Information Block (MIB). 
The UE then retrieves essential system information, including System Information Block~1 (SIB1), and monitors control resources such as Control Resource Set~\#0 (CORESET\#0) in NR before normal scheduling resumes. 
When uplink synchronization is required, the UE executes the Random Access Procedure (RACH) via the Physical Random Access Channel (PRACH). 
The establishment, maintenance, and release of radio connections are subsequently managed by the Radio Resource Control (RRC) protocol~\cite{3gpp38331}. 
Once an RRC connection is established, the UE operates in the RRC\_CONNECTED state.
When traffic becomes inactive, the UE may transition to RRC\_IDLE or, in NR deployments, optionally to RRC\_INACTIVE, which preserves connection context. 
The steering mechanisms considered in this work operate while the UE remains in RRC\_CONNECTED. 

\subsection{Steering Mechanisms}
\label{sec:runtime_ue}

Table~\ref{tab:steering_mechanism} summarizes mechanisms that steer RRC\_CONNECTED UEs across the spectrum.
To the best of our knowledge, this is the first unified UE-side PHY-layer characterization of steering mechanisms.
These mechanisms differ substantially in the PHY-layer procedures they trigger. 
BWP reconfiguration performs intra-cell bandwidth reconfiguration while preserving synchronization with the serving cell, resulting in minimal PHY-layer disruption. 
CA and EN-DC extend the available spectrum by activating additional SCells while maintaining the original control-plane anchor, requiring secondary-cell synchronization but avoiding full serving-cell migration.
In contrast, HO transfers control-plane responsibility to a target cell, requiring synchronization with the new cell before scheduling can resume. 
Finally, R\&R releases the connection, 
forcing full cell acquisition and RRC re-establishment. 
These differences imply distinct PHY-layer processing requirements that directly influence UE-side disruption during spectrum steering.

\begin{table}[b]
\centering
\caption{PHY-layer milestones from modem traces.}
\label{tab:latency_mapping}
\scriptsize
\renewcommand{\arraystretch}{1}
\setlength{\tabcolsep}{2pt}

\begin{tabular}{|p{2.1cm}|p{2.4cm}|p{3.8cm}|}
\hline
\textbf{Procedure} &
\textbf{Trigger ($T_0$)} &
\textbf{Observable Modem Milestones} \\
\hline

NR Baseline &
Initial connection &
SSB detection $\rightarrow$ PBCH/MIB decode $\rightarrow$ SIB1 acquisition $\rightarrow$ S-criteria success \\
\hline

LTE Baseline &
Initial connection &
PBCH/MIB decode $\rightarrow$ SIB1 acquisition $\rightarrow$ PDCCH decode $\rightarrow$ S-criteria success \\
\hline

BWP Reconfiguration &
RRCReconfiguration (BWP update) &
Configuration start $\rightarrow$ BWP configuration apply $\rightarrow$ Configuration completion \\
\hline

Carrier Aggregation &
RRCReconfiguration (SCell) &
SCell configuration $\rightarrow$ SCC configuration event $\rightarrow$ SCC activation $\rightarrow$ SCell measurements \\
\hline

EN-DC (NSA) &
RRCReconfiguration (NR Secondary Node) &
NR sync acquisition $\rightarrow$ NR RRC reconfiguration $\rightarrow$ NR carrier activation $\rightarrow$ NR measurement confirmation \\
\hline

Connected-Mode Handover &
RRCReconfiguration (mobilityControlInfo) &
Handover start $\rightarrow$ Target-cell synchronization $\rightarrow$ 
Scheduling resumes \\
\hline

Release and Redirection &
RRCRelease / RRCConnectionRelease (redirectedCarrierInfo) &
PBCH/MIB decode $\rightarrow$ SIB1 acquisition $\rightarrow$ S-criteria success  \\
\hline

\end{tabular}
\end{table}

\subsection{O-RAN Control Model}
\label{sec:o_ran_explained}

O-RAN introduces a disaggregated RAN architecture with a programmable control framework centered around the Near-Real-Time RIC (NearRT-RIC), operating on 10ms-1s timescales.
NearRT-RIC interfaces with distributed RAN nodes (E2 nodes) through the E2 interface and enables closed-loop control 
via xApps. 
Communication between NearRT-RIC and E2 nodes is governed by E2 Application Protocol (E2AP), which supports telemetry reporting. 
Specific functionalities are implemented by xApps through E2 Service Models (E2SMs), which are encapsulated within E2AP messages, allowing the exposure of data from E2 Nodes to the RIC.

\section{\system: System Overview}
\label{sec:system_overview}


In this section, we describe the PHY-layer to modem-trace mapping, latency decomposition, and disruption-aware steering policy underlying \system.

\begin{figure*}
    \centering
    \includegraphics[width=2\columnwidth]{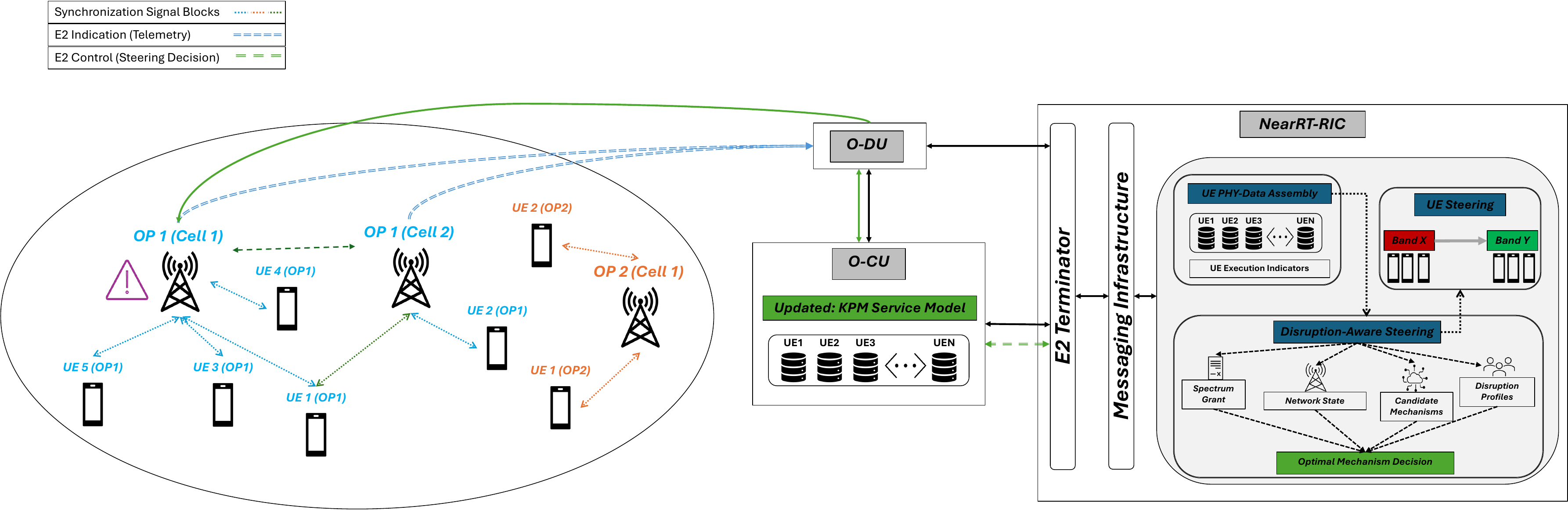}
    \caption{Overview of \system, illustrating modem trace extraction, latency characterization, and xApp-driven steering decisions.}
\label{fig:system_architecture}
\end{figure*}

\subsection{Modem-Level Event Extraction}

PHY-layer events are identified and extracted from UE-side modem traces obtained through vendor-supported diagnostic interfaces. 
Our analysis focuses on three logging layers that capture different stages of UE PHY-layer processing:

\begin{itemize}

\item \textit{Modem Layer 1 (ML1):} cell search, synchronization acquisition, cell measurements, and S-criteria evaluation during cell selection and camping.
\item \textit{Low Layer 1 (LL1):} control channel decoding and PHY-layer scheduling events.
\item \textit{Layer 2 (L2):} MAC-level configuration, activation procedures, and scheduling confirmations.

\end{itemize}

These logs expose the internal UE processing pipeline for each steering mechanism and, together with captured RRC messages, enable reconstruction of the PHY-layer milestone sequence summarized in 
Table~\ref{tab:latency_mapping}.
For completeness, we also include baseline NR and LTE initial connection procedures, as they reproduce several of the PHY-layer operations triggered by other mechanisms.
Steering initiation is anchored at the first control-plane trigger ($T_0$), extracted from the RRC message invoking the procedure. 
Completion ($T_F$) is determined using PHY-layer evidence of successful execution, such as scheduling continuity, cell measurement availability, or successful camping on the target resource, depending on the mechanism. 
Between $T_0$ and $T_F$, we reconstruct the sequence of intermediate modem milestones, enabling attribution of latency to individual PHY-layer stages.
The \textit{RRC-to-PHY} completion latency is defined as

\begin{equation}
T_{\text{RRC-PHY}} = T_F - T_0
\end{equation}

To isolate intrinsic modem execution cost from coordination-induced delay, we decompose $T_{\text{RRC-PHY}}$ into two components. 
Let $T_{\text{PHY}}$ denote the time elapsed between the first PHY-layer activation event and the final completion milestone $T_F$, this captures the \textit{PHY-centric} execution cost, reflecting the modem's intrinsic processing time independent of RRC signaling. 
The remaining interval, representing the modem's reaction delay to the RRC trigger, is expressed as

\begin{equation}
T_{\text{react}} = T_{\text{RRC-PHY}} - T_{\text{PHY}}
\end{equation}

\subsection{Disruption-Aware Steering Policy}

Different steering mechanisms introduce distinct levels of disruption, motivating steering decisions that account for the execution costs.
Let $\mathcal{M} = \{\text{BWP}, \text{CA}, \text{EN-DC}, \text{HO}, \text{R\&R}\}$ denote the set of steering mechanisms available. 
Whenever new spectrum resources become available, the RAN must select a mechanism $m \in \mathcal{M}$ to migrate traffic. 
Using the collected modem traces, we construct latency distributions, 
where $T^{(m)}$ denotes the execution latency of mechanism $m$.
The latency distribution $F_m(t)$ is defined as

\begin{equation}
F_m(t) = \Pr(T^{(m)} \leq t)
\end{equation}

From this distribution, we derive percentile-based disruption metrics that capture the variability of mechanism execution.
The $p$-th percentile latency is defined as

\begin{equation}
T_p^{(m)} = F_m^{-1}(p)
\end{equation}

where $p$ corresponds to high-percentile values that characterize worst-case disruption behavior.
To jointly capture intrinsic execution cost and RRC-to-PHY completion instability, we define a disruption score that combines PHY-centric latency with a multiplicative RRC-to-PHY completion variability term.
The disruption score is defined as


\begin{multline}
\label{eq:disruption_score}
D_m(\lambda, \mu) =
\Bigl[
\lambda \hat{T}_{95,\text{PHY}}^{(m)} +
(1-\lambda)\hat{\mathbb{E}}[T_{\text{PHY}}^{(m)}]
\Bigr] \\
\times \Bigl(1 + \mu \hat{V}_{\text{RRC-to-PHY}}^{(m)}\Bigr)
\end{multline}

where $\lambda \in [0,1]$ controls the tradeoff between tail and average PHY-centric execution latency and $\mu \in [0,1]$ controls the influence of RRC-to-PHY completion variability.
Smaller values of $\lambda$ minimize execution latency, while larger values prioritize avoiding worst-case events.
Larger $\mu$ values favor mechanisms with more predictable execution behavior, while $\mu = 0$ reduces the formulation to a purely PHY-centric score.
$\hat{V}_{\mathrm{RRC-to-PHY}}^{(m)}$ captures the relative variability introduced by modem reaction time, and we define it as

\begin{equation}
\hat{V}_{\mathrm{RRC-to-PHY}}^{(m)} =
\frac{\mathrm{IQR}_{\mathrm{RRC-to-PHY}}^{(m)}}{\tilde{T}_{\mathrm{RRC-to-PHY}}^{(m)}}
\end{equation}

where $\mathrm{IQR}_{\mathrm{RRC-to-PHY}}^{(m)}$ denotes the interquartile range of RRC-to-PHY completion latency and $\tilde{T}_{\mathrm{RRC-to-PHY}}^{(m)}$ its median, corresponding to the relative variability metric reported in Table~\ref{tab:derived_metrics}; all score components are normalized across mechanisms prior to evaluation.
This formulation 
balances PHY-centric efficiency with robustness to modem-reaction time.
Steering selection is therefore expressed as the disruption-aware decision rule

\begin{equation}
m^* = \arg\min_{m \in \mathcal{M}} D_m(\lambda, \mu)
\end{equation}

This decision logic is implemented as illustrated in Fig.~\ref{fig:system_architecture}.
UE modem traces are analyzed to extract PHY-layer milestones and construct disruption profiles for each mechanism, bootstrapped from an initial offline phase and periodically refreshed as new traces are gathered, ensuring disruption scores reflect current PHY-layer execution characteristics. 
During operation, the xApp receives network telemetry and spectrum availability information from E2 nodes via the 3GPP Key Performance Measurement (KPM) Service Model, extended by \system, and evaluates the disruption score for the available mechanisms. 
The mechanism minimizing the expected disruption is selected and executed via the corresponding E2 Control action to the RAN, triggering the appropriate steering procedure, as depicted in Fig.~\ref{fig:system_architecture}.

\section{Experimental Evaluation}
\label{sec:eval}


We evaluate \system\ using real-world measurements, covering experimental setup, mechanism characterization, and disruption profile derivation.

\subsection{O-RAN Compliant Testbed}

To validate the mapping presented in Table~\ref{tab:latency_mapping} and evaluate the disruption-aware steering policy within an O-RAN control loop, we deploy a controlled testbed. 
All software components run on a single x86\_64 host (Ubuntu 22.04.4 LTS, Intel i7-1195G7 CPU, 32\,GB RAM). 
The RAN functionality is provided by srsRAN~\cite{srsRAN}, while Open5GS~\cite{open5GS} implements the core network. 
FlexRIC~\cite{flexric} is used to provide NearRT-RIC support and E2 connectivity for closed-loop control. 
Ettus USRP X310 SDR devices serve as the RF front-end for the gNB/eNB, while multiple COTS UEs equipped with sysmocom SIMs attach to the network and execute the steering mechanisms under study.


\subsection{Real-World Data Collection}
\label{subsec:real-wolrd}

To characterize real-world PHY-layer performance, we collect measurements from commercial NR/LTE deployments across 12 urban areas in Boston and  San Francisco over a four-month period (2025-2026).
Data are captured on rooted COTS UEs (two Google Pixel~5 devices, one LG Velvet~5G, and one OnePlus~8~5G) using vendor-supported diagnostic logging and analyzed offline with Qualcomm QXDM/QCAT to extract synchronized ML1, LL1, L2, and RRC traces.
The resulting dataset contains 1,600 complete steering executions, yielding approximately 5,500 timestamped modem milestones. 
Among the observed mechanisms, HO appears most frequently ($n=574$), followed by BWP reconfiguration ($n=458$) and EN-DC ($n=302$). 
CA events occur less frequently ($n=49$), while R\&R executions are rare ($n=11$). 
The distribution of observations across mechanisms reflects real-world deployment conditions rather than collection bias and is corroborated by controlled testbed measurements.

\subsection{Real-World Performance Evaluation}
\label{subsec:performance_eval}

We characterize each mechanism under both PHY-centric and RRC-to-PHY completion timing, isolating intrinsic modem execution cost from coordination-induced reaction delays.


\begin{figure}[t]
    \centering
    \includegraphics[width=0.9\columnwidth]{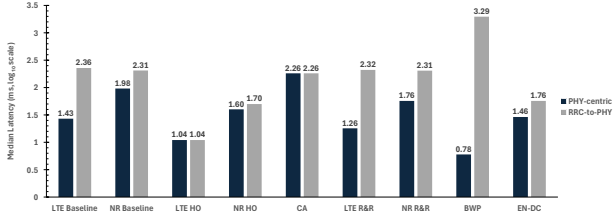}
    \caption{Median UE procedure latency (log scale).}
    \label{fig:median_latency_phy_vs_ota}
\end{figure}

\subsubsection{Median Disruption}

Fig.~\ref{fig:median_latency_phy_vs_ota} compares the median latency of the evaluated steering mechanisms under PHY-centric and RRC-to-PHY completion conditions. 
Under the PHY-centric view, BWP reconfiguration exhibits the smallest intrinsic disruption ($6.25\,\mathrm{ms}$), approximately $45\%$ faster than LTE HO, the second-fastest mechanism, and $85\%$ faster than NR HO. 
This behavior is expected, as BWP updates remain intra-cell and preserve synchronization with the serving cell, avoiding the additional synchronization procedures required by mobility mechanisms. 
Consequently, LTE HO appears faster than NR HO due to the additional complexity introduced by NR mobility procedures. 
A similar trend is observed during initial connection procedures. 
Both baseline connection events and R\&R exhibit the same technology-dependent behavior, with LTE remaining approximately $71\%$ and $68\%$ faster than NR, respectively, reflecting the higher acquisition complexity of NR cells. 
Finally, EN-DC demonstrates performance comparable to the LTE baseline ($29\,\mathrm{ms}$) and remains faster than the NR baseline, as many configuration parameters are pre-delivered through LTE RRC signaling.

However, when RRC-to-PHY completion timing is considered, the ranking changes substantially. 
BWP reconfiguration exhibits a significant increase to $1.97\,\mathrm{s}$ under RRC-to-PHY completion timing. 
This behavior indicates that coordinated PHY-layer reconfiguration within the modem cannot occur immediately, as BWP updates must preserve continuous service without interrupting ongoing transmissions. 
In contrast, HO procedures exhibit only modest amplification, since the mechanism already interrupts the PHY-layer processing pipeline, allowing modem procedures to begin immediately. 
For baseline connection procedures and R\&R, 
we observe faster modem reaction times for NR compared to LTE. 
This is due to NR’s shorter slot durations and faster control-to-PHY scheduling timelines, enabling the modem to initiate faster execution, even though LTE exhibits lower PHY-centric execution latency.
Finally, CA exhibits effectively zero reaction delay, as shown in Fig.~\ref{fig:median_latency_phy_vs_ota}, since it does not require full cell acquisition, allowing the transition to proceed with minimal signaling overhead. 

To understand the origin of per-mechanism amplification, we analyze the intermediate PHY-layer execution milestones. 
For most procedures, the dominant latency component is the interval between PBCH/MIB decoding and SIB1 acquisition, accounting for more than $80\%$ of the overall disruption and reaching up to $95\%$ in several cases. 
In contrast, the subsequent transition from SIB1 reception to camping is typically only on the order of $1$--$3\,\mathrm{ms}$. 
This result indicates that the dominant source of disruption is not UE processing but the waiting time for broadcast system information.

\begin{table}[b]
\centering
\caption{Amplification and variability metrics.}
\label{tab:derived_metrics}
\footnotesize
\setlength{\tabcolsep}{1pt}
\renewcommand{\arraystretch}{1}
\begin{tabular*}{\columnwidth}{@{\extracolsep{\fill}} l c c @{}}
\toprule
\textbf{Mechanism} & \textbf{RRC-to-PHY/PHY Ratio} & \textbf{Rel. Variability} \\
\midrule
LTE HO             & 1.0   & 0.63 \\
NR HO              & 1.3   & 0.46 \\
EN-DC                & 2.0   & 0.50 \\
\midrule
NR Baseline  & 5.1   & 0.69 \\
LTE Baseline & 8.4   & 0.94 \\
LTE R\&R             & 11.7  & 0.10 \\
NR R\&R              & 3.6   & 0.15 \\
\midrule
BWP                & 328.5 & 5.68 \\
CA                 & 0.9   & 3.15 \\
\bottomrule
\end{tabular*}
\end{table}

\begin{figure*}[t]
    \centering
    \subfloat[PHY-centric]{%
        \includegraphics[width=0.48\textwidth]{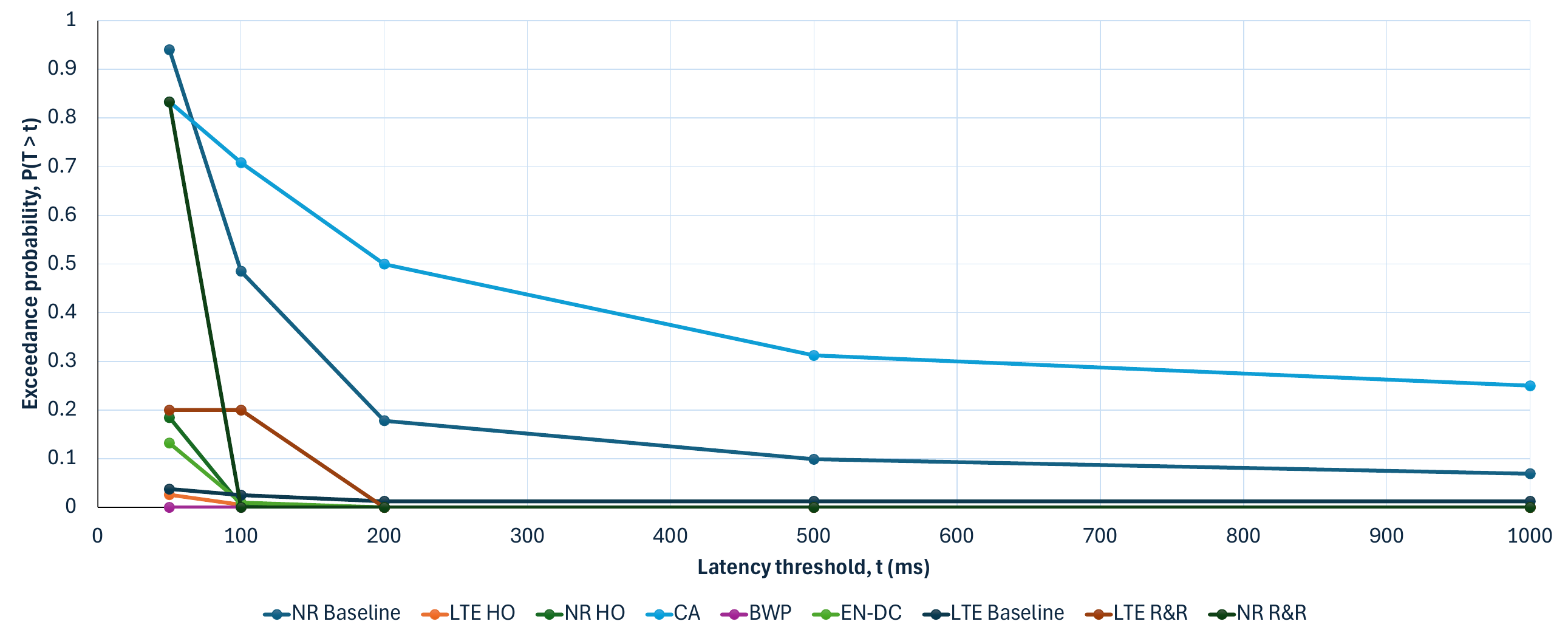}
        \label{fig:phy_tail}
    }
    \hfill
    \subfloat[RRC-to-PHY completion]{%
        \includegraphics[width=0.48\textwidth]{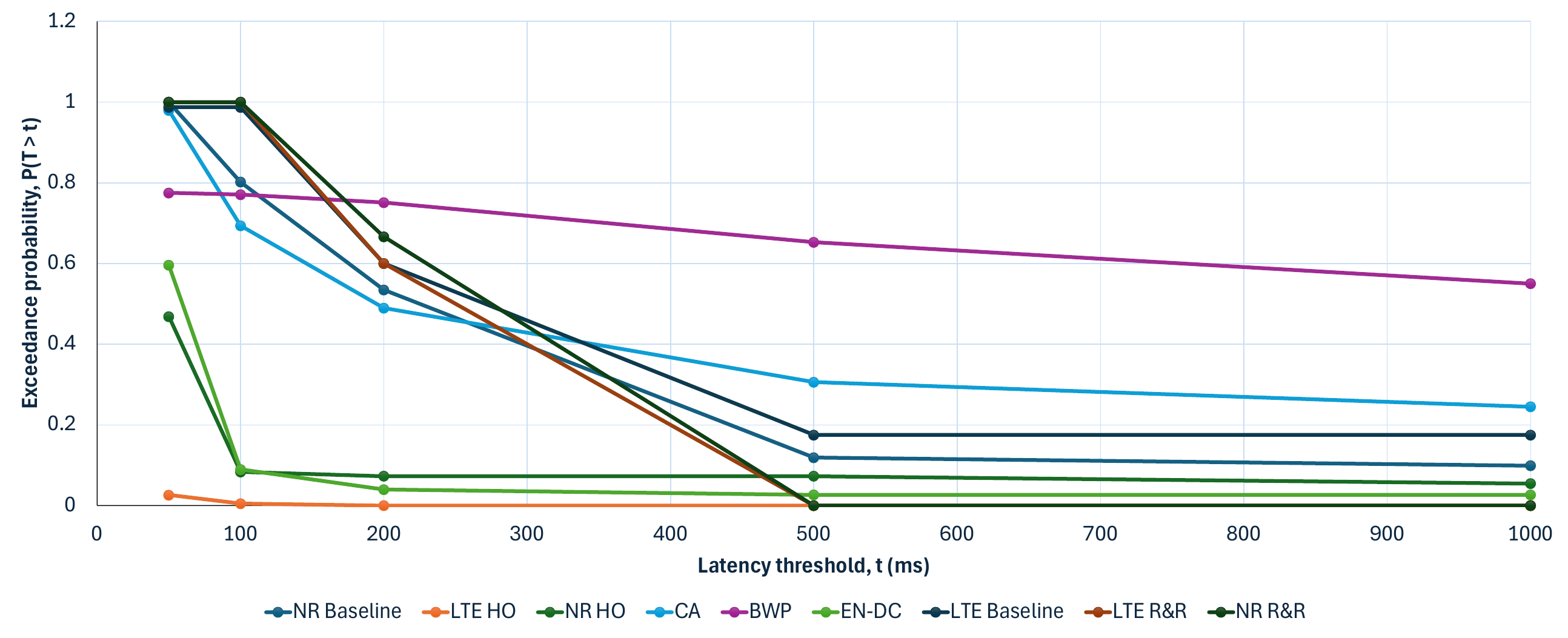}
        \label{fig:ota_tail}
    }
    \caption{Exceedance probability $P(T>t)$ across latency thresholds under PHY-centric and RRC-to-PHY completion timing.}
    \label{fig:tail_exceedance}
\end{figure*}

\subsubsection{Tail Behavior and Stability}

Median latency alone does not capture the stability of the analyzed mechanisms. 
In Fig.~\ref{fig:phy_tail} and~\ref{fig:ota_tail}, we therefore report the exceedance probability $P(T > t)$ across latency thresholds to characterize tail behavior. 
Under PHY-centric timing, most mechanisms exhibit rapidly decaying tails, with only baseline access and CA showing non-negligible probability of exceeding $200\,\mathrm{ms}$. 
This indicates that, in isolation, modem-side execution remains stable even for mechanisms with higher median latency, except for procedures involving new cell acquisition, particularly when the target cell has not been previously detected.
In contrast, RRC-to-PHY completion view reveals substantially heavier tails, highlighting significant amplification due to internal coordination delays. 
Fig.~\ref{fig:ota_tail} shows that this effect varies across mechanisms, with BWP reconfiguration exhibiting the most pronounced tail behavior, exceeding $1\,\mathrm{s}$ in more than $50\%$ of cases. 
These results demonstrate that RRC-to-PHY completion disruption is not solely determined by intrinsic UE processing but is strongly influenced by internal coordination overhead.

Table~\ref{tab:derived_metrics} summarizes these effects using two derived indicators: the RRC-to-PHY/PHY ratio $\tilde{T}_{\mathrm{RRC-to-PHY}}/\tilde{T}_{\mathrm{PHY}}$, capturing the amplification between RRC-to-PHY completion and PHY-centric views, and the relative variability, reflecting the stability of each mechanism under RRC-to-PHY completion timing. 
Mechanisms whose amplification remains close to $1$ behave as modem-bound procedures, where RRC-to-PHY completion latency closely tracks PHY-centric execution, as observed for HO and CA. 
Mechanisms involving cell discovery or system reconfiguration, however, exhibit significantly higher amplification factors due to signaling delays and coordination overhead. 
Regarding stability, most mechanisms, including baseline access, HO, and EN-DC, exhibit relatively low variability, with $\hat{V}_{\mathrm{RRC-to-PHY}}^{(m)} < 1$. 
Conversely, CA and BWP show substantially higher variability, as also observed in Fig.~\ref{fig:ota_tail}. 
For BWP, this variability arises from delayed execution of coordinated PHY-layer reconfiguration, while for CA, it reflects variability in configuration of PHY-centric procedures.

\subsection{Disruption-Aware Steering Policy Evaluation}
\label{subsec:xAPP_performance}

To evaluate \system\, we apply Eq.~\ref{eq:disruption_score}.    
We sweep $\lambda \in \{0, 0.25, 0.5, 0.75, 1\}$ and $\mu \in \{0, 0.25, 0.5, 0.75, 1\}$ and compare \system\ against static baselines, including always-BWP, always-HO, min-mean, and min-$T_{95}$, across five scenarios: 
unconstrained, no-BWP, mobility-only, LTE-only, and HO-or-BWP.
In the unconstrained and HO-or-BWP scenarios, \system\ consistently selects BWP 
achieving 
zero exceedance above $50\,\mathrm{ms}$, and reducing mean latency by $85.1\%$ and $T_{95}$ by $89.7\%$  over always-HO, while matching min-mean and min-$T_{95}$.
When BWP is unavailable, always-BWP fails by definition, while \system\ adapts and selects the least disruptive available mechanism, reducing mean latency by $64.2\%$ and $T_{95}$ by $65.4\%$ over always-HO, and correctly avoiding CA, whose mean latency of $1225\,\mathrm{ms}$ would represent a $98.8\%$ degradation relative to the selected mechanism.
In the mobility-only and LTE-only scenarios, \system\ additionally avoids R\&R variants, which incur up to $74.8\%$ higher $T_{95}$ relative to the selected mechanism.
Selection remains optimal,
while adapting across scenarios, confirming robustness to parameter settings. 
Across all scenarios, \system\ achieves consistent reductions of up to $85.1\%$ in mean latency, $89.7\%$ in $T_{95}$, and full elimination of tail exceedance above $50\,\mathrm{ms}$ relative to operator-default static policies, demonstrating that PHY-centric disruption profiling enables reliable, context-aware mechanism selection.

\section{Conclusions}
\label{sec:conclusions}


We present the first PHY-aware characterization of 3GPP-compliant UE steering mechanisms for DSS, showing that similar mechanisms impose fundamentally different UE-side disruption costs, with NR BWP achieving $6.25\,\mathrm{ms}$ mean latency and CA exceeding $1225\,\mathrm{ms}$.
Building on modem-trace decomposition into PHY-centric and RRC-to-PHY completion components, we design \system, a NearRT-RIC-based system that selects the least disruptive mechanism via a two-parameter disruption score.
Across five deployment scenarios, \system\ reduces mean latency by up to $85.1\%$ and $T_{95}$ by $89.7\%$ over static policies, fully eliminates tail exceedance above $50\,\mathrm{ms}$, and avoids mechanisms such as CA that would degrade mean latency by $98.8\%$.

\balance
\bibliographystyle{unsrt}
\bibliography{ref}

\end{document}